\documentclass[12pt]{article}
\usepackage[dvips]{epsfig}

\begin{document}

\author{A. de Souza Dutra$^{a,b}$\thanks{%
E-mail: dutra@feg.unesp.br} \\
$^{a}$Abdus Salam ICTP, Strada Costiera 11, 34014 Trieste Italy\\
$^{b}$UNESP - Campus de Guaratinguet\'{a} - DFQ\thanks{%
Permanent Institution}\\
Av. Dr. Ariberto Pereira da Cunha, 333\\
C.P. 205\\
12516-410 Guaratinguet\'{a} SP Brasil}
\title{{\LARGE General solutions for some classes of interacting two field kinks}}
\maketitle

\begin{abstract}
In this work we present some classes of models whose the corresponding two
coupled first-order nonlinear equations can be put into a linear form, and
consequently be solved completely. In these cases the so called trial orbit
method is completely unnecessary. We recall that some physically important
models as, for instance, the problem of tiling a plane with a network of
defects and polymer properties are in this class of models.

PACS numbers: 11.27.+d, 11.30.Er
\end{abstract}

\newpage

A rapid look at the history of physics is enough to lead anyone to conclude
that, fortunately, the most part of the natural physical systems can be
studied by using linear differential equations, with their good properties
like the superposition principle. Notwithstanding, there are some classes of
important systems with are intrinsically nonlinear and, nowadays, there is a
growing interest in dealing with such systems \cite{witten} - \cite{fosco}.
Unfortunately, as a consequence of the nonlinearity, in general we lose the
capability of getting the complete solutions. In this work we show that for
those systems in 1+1 dimensions, whose the second-order differential
equations can be reduced to the solution of corresponding first-order
equations, the so called Bolgomol 'nyi-Prasad-Sommerfield (BPS) topological
solitons \cite{BPS}, one can obtain a differential equation relating the two
coupled fields which, once solved, leads to the general orbit connecting the
vacua of the model. In fact, the ``trial and error '' methods historically
arose as a consequence of the intrinsic difficulty of getting general
methods of solution for nonlinear differential equations. About two decades
ago, Rajaraman \cite{rajaraman} introduced an approach of this nature for
the treatment of coupled relativistic scalar field theories in 1+1
dimensions. His procedure was model independent and could be used for the
search of solutions in arbitrary coupled scalar models in 1+1 dimensions.
However, the method is limited in terms of the generality of the solutions
obtained and is convenient and profitable only for some particular, but
important, cases \cite{boya}. Some years later, Bazeia and collaborators
\cite{bazeia0} applied the approach developed by Rajaraman to special cases
where the solution of the nonlinear second-order differential equations are
equivalent to the solution of corresponding first-order nonlinear coupled
differential equations. By the way, Bazeia and collaborators wisely applied
their solution to a variety of natural systems, since polymers up to domain
walls. In this work we are going to present a procedure which is absolutely
general when applied to systems like those described in \cite{bazeia0},
namely the BPS topological solutions. Furthermore, we are going also to show
that many of the systems studied in \cite{bazeia0}-\cite{bazeia4} can be
mapped into a first-order linear differential equation and, as a
consequence, can be solved in order to get the general solution of the
system. After that, we trace some comments about the consequences coming
from these general solutions.

In order to deal with the problem, following the usual procedure to get BPS
\cite{BPS} solutions for nonlinear systems, one can particularize the form
of the Lagrangian density
\begin{equation}
L=\frac{1}{2}\left( \partial _{\mu }\phi \right) ^{2}+\frac{1}{2}\left(
\partial _{\mu }\chi \right) ^{2}-V\left( \phi ,\chi \right) ,
\end{equation}

\noindent by imposing that the potential must be written in terms of a
superpotential like
\begin{equation}
V\left( \phi ,\chi \right) =\frac{1}{2}\left( \frac{\partial W\left( \phi
,\chi \right) }{\partial \phi }\right) ^{2}+\frac{1}{2}\left( \frac{\partial
W\left( \phi ,\chi \right) }{\partial \chi }\right) ^{2}.
\end{equation}

The energy of the so-called BPS states can be calculated straightforwardly,
giving
\begin{equation}
E_{B}=\frac{1}{2}\int_{-\infty }^{\infty }dx\left[ \left( \frac{d\phi }{dx}%
\right) ^{2}+\left( \frac{d\chi }{dx}\right) ^{2}+\,W_{\phi }^{2}+\,W_{\chi
}^{2}\right] ,
\end{equation}

\noindent which lead us to
\begin{equation}
E_{B}=\frac{1}{2}\int_{-\infty }^{\infty }dx\left[ \left( \frac{d\phi }{dx}%
-W_{\phi }\right) ^{2}+\left( \frac{d\chi }{dx}-W_{\chi }\right)
^{2}+2 \,\left( W_{\chi }\frac{d\chi }{dx}+W_{\phi }\frac{d\phi
}{dx} \right) \right] ,
\end{equation}

\noindent and finally to

\begin{equation}
E_{B}=|W\left( \phi _{j},\chi _{j}\right) -W\left( \phi _{i},\chi
_{i}\right) |,  \label{eq3}
\end{equation}

\noindent where $\phi _{i}$ and $\chi _{i}$ are the are the $i-th$ vacuum
state of the model \cite{bazeia1.5}.

In this case, one can easily see that solutions with minimal energy of the
second-order differential equation for the static solutions in 1+1
dimensions, can be solved through the corresponding first-order coupled
nonlinear equations
\begin{equation}
\frac{d\phi }{dx}=W_{\phi }\left( \phi ,\chi \right) \,,\,\frac{d\chi }{dx}%
=W_{\chi }\left( \phi ,\chi \right) ,  \label{eq1}
\end{equation}

\noindent where $W_{\phi }\equiv \frac{\partial W}{\partial \phi }$ and $%
W_{\chi }\equiv \frac{\partial W}{\partial \chi }$. Here, it is
important to remark that the BPS solutions settle into vacuum
states asymptotically. In other words, the vaccum states act as
implicit boundary conditions of the BPS equations.

Now, instead of applying the usual trial-orbit approach \cite{bazeia0}-\cite
{bazeia4}, we note that it is possible to write the following equation
\begin{equation}
\frac{d\phi }{W_{\phi }}=dx=\frac{d\chi }{W_{\chi }},
\end{equation}

\noindent where the spatial differential element is a kind of invariant. So,
one obtains that
\begin{equation}
\frac{d\phi }{d\chi }=\frac{W_{\phi }}{W_{\chi }}.  \label{eqm}
\end{equation}

This last equation is, in general, a nonlinear differential equation
relating the scalar fields of the model. Now, if one is able to solve it
completely, the function $\phi \left( \chi \right) $ can be used to
eliminate one of the fields, so rendering the equations (\ref{eq1})
uncoupled. Finally, this uncoupled first-order nonlinear equation can be
solved in general, even if numerically.

From now on, we choose a particular model which can be used for modelling a
number of systems \cite{bazeia1.5}, in order to exemplify the method in a
concrete situation. In fact we will show that for this situation, the
equation (\ref{eqm}) can be mapped into a linear differential equation, from
which it is possible to obtain the general solution. In this case the
superpotential is written as
\begin{equation}
W\left( \phi ,\chi \right) =-\lambda \,\phi +\frac{\lambda }{3}\phi ^{3}+\mu
\,\phi \,\chi ^{2},  \label{w1}
\end{equation}

\noindent and the equation (\ref{eqm}) looks like
\begin{equation}
\frac{d\phi }{d\chi }=\frac{\lambda \left( \phi ^{2}-1\right) +\mu \,\chi
^{2}}{2\,\mu \,\phi \,\chi }.
\end{equation}

At this point one can verify that, performing the transformation $\phi
^{2}=\rho +1$. The above equation can be written as
\begin{equation}
\frac{d\rho }{d\chi }-\frac{\lambda }{\mu \,\chi }\,\rho =\chi ,
\end{equation}

\noindent a typical inhomogeneous linear differential equation. It
is interesting to observe that its particular solution,
corresponds to the result usually presented in the literature
\cite{bazeia1.5}. The general solution is easily obtained, giving
\begin{equation}
\rho \left( \chi \right) =\phi ^{2}-1=c_{0}\,\chi ^{\frac{\lambda }{\mu }}-%
\frac{\mu }{\lambda -2\,\mu }\,\,\chi ^{2},
\end{equation}

\noindent for $\lambda \neq 2\,\mu $, and
\begin{equation}
\phi ^{2}-1=\chi ^{2}\left[ \ln \left( \chi \right) +c_{1}\right] ,
\end{equation}

\noindent for the $\lambda =2\,\mu $ case, and $c_{0}$ and $c_{1}$
are arbitrary integration constants. It is interesting to note the
this last particular situation was not taken into consideration in
the literature up to our knowledge. From now on, we substitute
these solutions in one of the equations (\ref{eq1}), and solve it,
so obtaining a generalized solution for the system. In general it
is not possible to solve $\chi $ in terms of $\phi $ from the
above solutions, but the contrary is always granted. Here we will
substitute $\phi \left( \chi \right) $ in the equation for the field $\chi $%
, obtaining
\begin{equation}
\frac{d\chi }{dx}=\pm \,2\,\mu \,\chi \,\sqrt{1+c_{0}\,\chi ^{\frac{\lambda
}{\mu }}-\left( \frac{\mu }{\lambda -2\,\mu }\right) \chi ^{2}},\,\,\left(
\lambda \neq 2\,\mu \right) ,  \label{eq2}
\end{equation}

\noindent and
\begin{equation}
\frac{d\chi }{dx}=\pm \,2\,\mu \,\chi \,\sqrt{1+\chi ^{2}\left[ \ln \left(
\chi \right) +c_{1}\right] },\,\left( \lambda =2\,\mu \right) .
\end{equation}

In general we can not have an explicit solution for the above equations.
However one can verify numerically that the solutions are always of the same
classes. Notwithstanding, some classes of solutions can be written in closed
explicit forms. First of all, we should treat the system when $c_{0}=0$,
because in this situation we can solve analytically the system for any value
of $\lambda $, apart from the case $\lambda =2\,\mu $. In this situation we
get
\begin{equation}
\chi _{+}\left( x\right) =\frac{2\,e^{2\,\mu \,\left( x-x_{0}\right) }}{%
1-c\,e^{4\,\mu \left( x-x_{0}\right) }},\,\,\chi _{-}\left( x\right) =\frac{%
2\,\,e^{4\,\mu \left( x-x_{0}\right) }}{c-\,e^{4\,\mu \left( x-x_{0}\right) }%
},
\end{equation}

\noindent with $c\equiv -\frac{\mu }{\lambda -2\,\mu }\,$. For this choice
of the parameters, the solution always vanishes at the boundary ($%
x\rightarrow \pm \infty $). As a consequence, the corresponding kink
solution for the field $\phi $, will be given by
\begin{equation}
\phi _{+}\left( x\right) =\pm \,\frac{c\,e^{4\,\mu \,\left( x-x_{0}\right)
}+1}{c\,e^{4\,\mu \,\left( x-x_{0}\right) }-1},\,\,\phi _{-}\left( x\right)
=\pm \,\frac{c+\,e^{4\,\mu \left( x-x_{0}\right) }}{c-\,e^{4\,\mu \left(
x-x_{0}\right) }},
\end{equation}

\noindent which are essentially equivalent to those solutions appearing in
\cite{bazeia1.5}, given in terms of $\tanh \left( x\right) $. Let us now
discuss below two particular cases $\left( c_{0}\,\neq 0\right) $ where the
integration can be performed analytically up to the end. Let us first
consider the case were $\lambda =\mu $, which has as solutions
\begin{equation}
\chi _{+}\left( x\right) =\frac{4\,e^{2\mu \left( x-x_{0}\right) }}{\left[
c_{0}\,e^{2\mu \left( x-x_{0}\right) }-1\right] ^{2}-4\,e^{4\mu \left(
x-x_{0}\right) }};\,\,\chi _{-}\left( x\right) =\frac{4\,e^{2\mu \left(
x-x_{0}\right) }}{\left[ \,e^{2\mu \left( x-x_{0}\right) }-c_{0}\right]
^{2}-4};
\end{equation}

\noindent where we must impose that $c_{0}\,\leq -\,2$ in both solutions, in
order to avoid singularities of the field as can be easily verified.
Furthermore, both solutions vanishes when $x\rightarrow \pm \infty $,
provided that $c_{0}\,\neq -\,2$. On the other hand the corresponding
solutions for the field $\phi \left( x\right) $ are given by
\begin{equation}
\phi _{+}\left( x\right) =\frac{\left( c_{0}^{2}-4\right) e^{4\mu \left(
x-x_{0}\right) }-1}{\left[ c_{0}\,e^{2\mu \left( x-x_{0}\right) }-1\right]
^{2}-4\,e^{4\mu \left( x-x_{0}\right) }};\,\phi _{-}\left( x\right) =\frac{%
4-c_{0}^{2}+e^{4\mu \left( x-x_{0}\right) }}{\left[ e^{2\mu \left(
x-x_{0}\right) }-c_{0}\right] ^{2}-4\,}.
\end{equation}

Here the first bonus coming from the complete exact solution of
the equation (\ref{eq1}) comes when we deal with the special case
with $c_{0}=-2$. It is remarkable that for this precise value of
the arbitrary integration constant, an absolutely unexpected kink
solution do appears. In fact, it could never be obtained from the
usually used solution, where $c_{0}=0$ necessarily. In this
special solution, the field $\chi $ is a kink with the following
asymptotic limits: $\chi _{+}\left( -\infty \right) =0$ and $\chi
_{+}\left( \infty \right) =1$, and $\phi _{+}\left( -\infty
\right) =-1$ and $\phi _{+}\left( \infty \right) =0$, and
correspondingly $\chi _{-}\left( -\infty \right) =1$ and $\chi
_{-}\left( \infty \right) =0$, and $\phi _{-}\left( -\infty
\right) =0$ and $\phi _{-}\left( \infty \right) =1$, as it can be
seen from an example of a typical profile of this kink in the Fig.
1 . Below we present a plot of this kink, which we are going to
call type B kink, in contrast with the other cases where the field
$\chi $ does not have a kink profile, which we call type A kink
(see Fig. 2) . An interesting observation is that the choice
$c_{0}=-2$, is precisely the  one which makes the right hand side
of equation (\ref{eq2}) simply proportional to $\chi \left| 1-\chi
\right| =\zeta \,\chi \left( 1-\chi \right) $, where $\zeta $ is
the sign function defined as $\zeta \equiv \left( 1-\chi \right)
/\left| 1-\chi \right| $. It  takes values $\pm 1$ with $\zeta
=+1\,$being selected by boundary conditions $0\leq \chi \leq 1$
for the solutions appearing in (18) and, in this situation, the
equation is much easier to solve. In fact, by performing the
translation $\chi=\beta+\frac{1}{2}$, we recover a BPS
superpotential for the ``$\lambda \, \phi^4 "$ model,
$-(\beta^2-1/4)$. A similar situation will happens with the next
example.

As the third particular case, we consider the situation where $\lambda =4\mu
$. Now, the exact solutions look like
\begin{eqnarray}
\chi _{+}\left( x\right) &=&-\,\frac{2\,e^{2\mu \left( x-x_{0}\right) }}{%
\sqrt{\left[ \frac{1}{2}\,e^{4\mu \left( x-x_{0}\right) }+1\right]
^{2}-4\,c_{0}\,\,e^{8\mu \left( x-x_{0}\right) }}};\,  \nonumber \\
&& \\
\,\chi _{-}\left( x\right) &=&-\,\frac{4\,\,e^{2\mu \left( x-x_{0}\right) }}{%
\sqrt{\left[ 1+2\,\,e^{4\mu \left( x-x_{0}\right) }\right] ^{2}-16\,c_{0}}};
\nonumber
\end{eqnarray}

\noindent which have the same asymptotic behavior as that presented in the
previous cases for the type A kinks. In other words, provided that $%
c_{0}\neq 1/16$, only the field $\phi $ will be a kink.
Afterwards, as in the previous case, if one wish to avoid
intermediary singularities, one must impose that $c_{0}\leq
\frac{1}{16}$. Now, the $\phi $ solutions will be written as
\begin{eqnarray}
\phi _{+}\left( x\right) &=&\frac{4+\left( 16\,c_{0}-1\right) \,e^{8\,\mu
\left( x-x_{0}\right) }}{\left[ 2+\,e^{4\,\mu \left( x-x_{0}\right) }\right]
^{2}-16\,c_{0}\,\,e^{8\,\mu \left( x-x_{0}\right) }}\,\,\,;  \nonumber \\
&& \\
\,\phi _{-}\left( x\right) &=& \,\frac{16\,c_{0}+4\,e^{8\,\mu \left(
x-x_{0}\right) }-1}{\left[ 1+2\;e^{4\,\mu \left( x-x_{0}\right) }\right]
^{2}-16\,c_{0}}.  \nonumber
\end{eqnarray}

Once more, the particular choice of the integration parameter $c_{0}=\frac{1%
}{16}$, generates a type B kink, with the asymptotic behavior given by: $%
\chi _{+}\left( -\infty \right) =0$ and $\chi _{+}\left( \infty \right) =-2$%
, and $\phi _{+}\left( -\infty \right) =1$ and $\phi _{+}\left( \infty
\right) =0$, and correspondingly $\chi _{-}\left( -\infty \right) =-2$ and $%
\chi _{-}\left( \infty \right) =0$, and $\phi _{-}\left( -\infty \right) =0$
and $\phi _{-}\left( \infty \right) =1$.

It is interesting to calculate the energy of these two species of solitonic
configurations. For this we use the superpotential (\ref{w1}) and substitute
it in the equation (\ref{eq3}), and observe that the type A kinks have an
energy given by $E_{A}^{BPS}=\frac{4}{3}\,\lambda $ and in the two cases
considered above ($\lambda =\mu $ and $\lambda =4\,\mu $) we
obtain\thinspace $E_{B}^{BPS}=\frac{2}{3}\,\lambda $. One could interpret
these solutions as representing two kinds of torsion in a chain, represented
through an orthogonal set of coordinates $\phi $ and $\chi $. So that, in
the plane ($\phi $,$\chi $), the type A kink corresponds to a complete
torsion going from $(-1,0)$ to $(0,0)$ and the type B corresponds to a half
torsion, where the system goes from $(-1,0)$ to $(0,1)$, in the case where ($%
\lambda =\mu $) for instance.

In what follows, we will study a more general model, contemplating a number
of particular cases which have been studied in the literature, including the
previous and some other new ones. For this, we begin by defining the
superpotential
\begin{equation}
W\left( \phi ,\chi \right) =\frac{\mu }{2}\,\phi ^{N}\,\chi ^{2}+G\left(
\phi \right) ,
\end{equation}

\noindent which lead us to the following set of equations:
\begin{equation}
\frac{d\phi }{dx}=\frac{dG\left( \phi \right) }{d\phi }+\frac{\mu }{2}%
\,N\,\,\phi ^{\left( N-1\right) }\,\chi ^{2};\,\,\,\,\frac{d\chi }{dx}=\mu
\,\phi ^{N}\,\chi .
\end{equation}

\noindent So, the corresponding equation for the dependence of the field $%
\phi $ as a function of the field $\chi $, is given by
\begin{equation}
\frac{d\phi }{d\chi }=\frac{\frac{dG\left( \phi \right) }{d\phi }+\frac{\mu
}{2}\,N\,\,\phi ^{\left( N-1\right) }\,\chi ^{2}}{\mu \,\phi ^{N}\,\chi }.
\end{equation}

Now, performing the transformation\thinspace $\sigma \equiv \phi ^{2}$ we
get
\begin{equation}
\frac{d\sigma }{d\chi }=N\,\,\chi +\frac{2\,G_{\phi }\left( \sigma \right) }{%
\mu \,\sigma ^{\left( \frac{N-1}{2}\right) }}\,\,\frac{1}{\chi },
\end{equation}

\noindent where $G_{\phi }\left( \sigma \right) \equiv \frac{dG\left( \phi
\right) }{d\phi }\mid _{\sigma =\phi ^{2}}$. Obviously, there are no
arbitrary solutions for the above equation, but for that ones with exact
solution we can get the corresponding exact two-field solitons. For
instance, let us treat the special case where
\begin{equation}
G_{\phi }\left( \sigma \right) \equiv \frac{dG\left( \phi \right) }{d\phi }%
\mid _{\sigma =\phi ^{2}}\,=\frac{2\,\left( a_{0}+a_{1}\,\sigma
+a_{2}\,\sigma ^{2}\right) }{\mu }\,\sigma ^{\frac{N-1}{2}}.
\end{equation}

\noindent The solution will be given by a combination of Bessel functions
which, once substituted in the equation for the field $\chi $, lead us to a
hardly exactly solvable equation, beyond some singularities which appear in
the solution. So, we still here continue to work with the simpler linear
case of this equation, where $a_{2}=0$, which furthermore permits us to
write arbitrary solutions given by
\begin{equation}
\sigma \left( \chi \right) =-\frac{a_{0}}{a_{1}}-\frac{N\,\mu \,\chi ^{2}}{%
2\,\left( \mu -a_{1}\right) }+c_{I}\,\,\chi ^{\frac{2\,a_{1}}{\mu }},
\end{equation}

\noindent with $G\left( \phi \right) $ given by
\begin{equation}
G\left( \phi \right) =\frac{\mu \,\,\phi ^{N}}{2}\left[ \frac{a_{0}}{N}+%
\frac{a_{1}}{\left( N+2\right) }\,\phi ^{2}\right] ,
\end{equation}

\noindent leaving us with the following potential
\begin{eqnarray}
V(\phi ,\chi ) &=&\frac{1}{2}\,\phi ^{2\left( N-1\right) }a_{1}^{2}\,\phi
^{4}+2\,a_{1}N\,\mu \,\phi ^{2}\chi ^{2}+a_{0}^{2}+\mu ^{2}\chi ^{2}\left(
4\,\phi ^{2}+N^{2}\chi ^{2}\right) +  \nonumber \\
&&+2\,a_{0}\left( a_{1}\phi ^{2}+N\,\mu \,\chi ^{2}\right) ,
\end{eqnarray}

\noindent with $c_{I}$ being the integration arbitrary constant, and $a_{0}$
and $a_{1}$ are constants which characterize the physical system. From
above, it is easy to conclude that
\begin{equation}
\phi =\pm \sqrt{-\frac{a_{0}}{a_{1}}-\frac{N\,\mu \,\chi ^{2}}{2\left( \mu
-a_{1}\right) }+c_{I}\,\,\chi ^{\frac{2\,a_{1}}{\mu }}},
\end{equation}

\noindent and, consequently we are left to solve the following equation
\begin{equation}
\frac{d\chi }{dx}=\pm \,\mu \,\left[ -\frac{a_{0}}{a_{1}}-\frac{N\,\mu
\,\chi ^{2}}{2\left( \mu -a_{1}\right) }+c_{I}\,\,\chi ^{\frac{2\,a_{1}}{\mu
}}\right] ^{\frac{N}{2}}\,\chi .
\end{equation}

At this point it is important to remark that many models appearing in the
literature can be cast as particular cases from the above general one. For
instance if we take $N=1$, we recover the models I, II and III of \cite
{bazeia1.5}, and model I of \cite{bazeia0}. The case where $N=2$ is
equivalent to the model II in \cite{bazeia0} and the model considered in
\cite{bazeia3}.

As a final comment we should say that one can even make a bit generalization
of the above exactly solved two fields models. This could be done by
starting with the superpotential
\begin{equation}
W_{NM}\left( \phi ,\chi \right) \equiv G\left( \phi \right) +\frac{\mu }{M}%
\,\phi ^{N}\,\chi ^{M},
\end{equation}

\noindent with $G\left( \phi \right) $ being the same appearing previously
in the text. After manipulations similar to that one done above, we end with
the equation
\begin{equation}
\frac{d\sigma \left( \chi \right) }{d\chi }=\left( \frac{2\,a_{1}}{\mu }%
\right) \sigma \left( \chi \right) \,\chi ^{\left( 1-M\right) }+\left( \frac{%
2\,N}{M}\right) \,\chi \,,
\end{equation}

\noindent where $\sigma \equiv \phi ^{2}+\left( \frac{a_{0}}{a_{1}}\right) $%
. Solving the above equation for arbitrary $M$, one obtains that
\begin{eqnarray}
\sigma _{M}\left( \chi \right) &=&e^{-\left[ \frac{2\,a_{1}\chi ^{\left(
2-M\right) }}{\mu \left( M-2\right) }\right] }\left\{ c_{1}+\frac{1}{M\left(
M-2\right) }\left[ 2^{\frac{M}{M-2}}\,\,N\,\,\,\chi ^{2}\Gamma \left( -\frac{%
2}{M-2},\right. \right. \right.  \nonumber \\
&& \\
&&\left. \left. -\frac{2\,a_{1}\chi ^{\left( 2-M\right) }}{\mu \left(
2-M\right) }\right) \right] \left. \left( \frac{a_{1}\chi ^{\left(
2-M\right) }}{\mu \left( 2-M\right) }\right) ^{\left( \frac{2}{M-2}\right)
}\right\} .  \nonumber
\end{eqnarray}

\noindent where $c_{1}$ is the arbitrary integration constant, and $\Gamma
\left( a,z\right) =\int_{z}^{\infty }t^{\left( a-1\right) }e^{-t}dt$, is the
incomplete gamma function.

Obviously, the case studied earlier in this work is obtained from the above
when one chooses $M=2$. On the other hand, we can get simpler solutions for
other particular values of the parameter M as, for instance $M=4$, whose
solution can be written as
\begin{equation}
\sigma _{4}\left( \chi \right) =\frac{N}{4}\,\chi ^{2}+e^{-\frac{a_{1}}{\mu
\chi ^{2}}}\left[ c_{1}+\frac{N\,a_{1}}{4\,\mu }\, Ei \left( \frac{a_{1}%
}{\mu \,\chi ^{2}}\right) \right] ,
\end{equation}

\noindent where $ Ei \left( z\right) \equiv -\,\int_{-z}^{\infty }$ $%
\frac{e^{-t}}{t}\,dt$, is the exponential integral function. It can be seen
from Fig.3 that, apart from a small region close to the origin, it is
asymptotically similar to that of the case with $M=2$, which was discussed
in some detail above in the text. This expression does not have any kind of
singularity and approaches to zero when the field $\chi $ does the same.
Notwithstanding, the last part of the analysis of the kinks needs to be done
through evaluation of the equation
\begin{equation}
\frac{d\chi }{dx}=\mu \,\chi ^{\left( M-1\right) }\,\,\left( \pm \right)
^{N}\,\left[ -\left( \frac{a_{0}}{a_{1}}\right) +\sigma _{M}\left( \chi
\right) \right] ^{\frac{N}{2}},
\end{equation}

\noindent which is not easy to be done analytically, so that one needs to
make use of numerical techniques. We intend to perform this analysis in a
future work, looking for new interesting features.

\vfill

\noindent \textbf{Acknowledgments:} The author is grateful to CNPq
for partial financial support, to the Professor D. Bazeia for
introducing him to this matter, and to the referee for the very
pertinent and constructive criticisms and suggestions. This work
has been finished during a visit within the Associate Scheme of
the Abdus Salam ICTP.

\newpage

\begin{figure}[tbp]
\begin{center}
\begin{minipage}{20\linewidth}
\epsfig{file=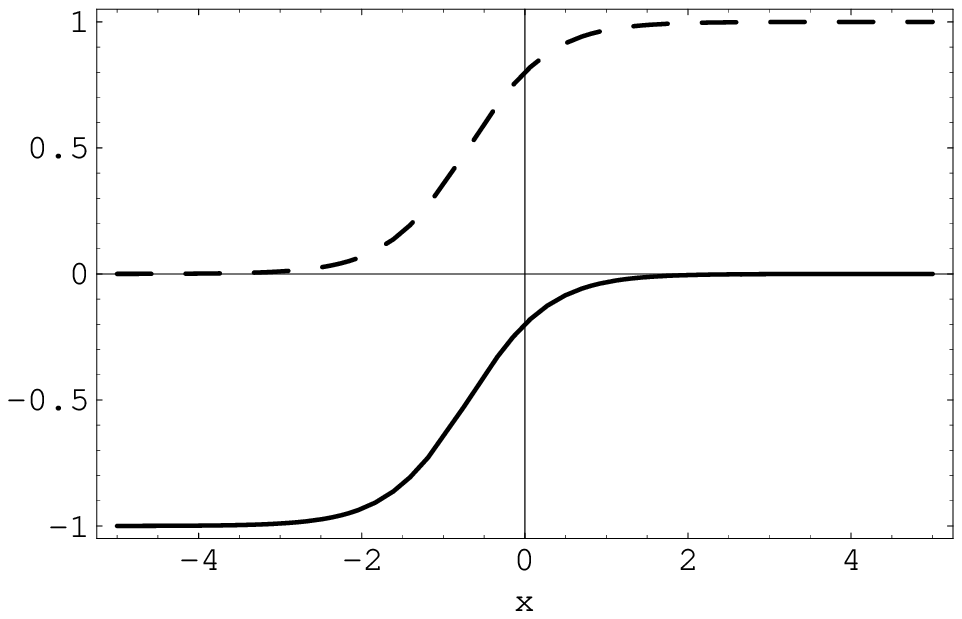}
\end{minipage}
\end{center}
\caption{Typical type B kink profile (for $\lambda=\mu$ ). The dotted line
corresponds to the field $\chi_+(x)$ and the solid line to the field $%
\phi_+(x)$. Both were calculated for $c_0=-2$. }
\label{fig:fig1}
\end{figure}

\begin{figure}[tbp]
\begin{center}
\begin{minipage}{20\linewidth}
\epsfig{file=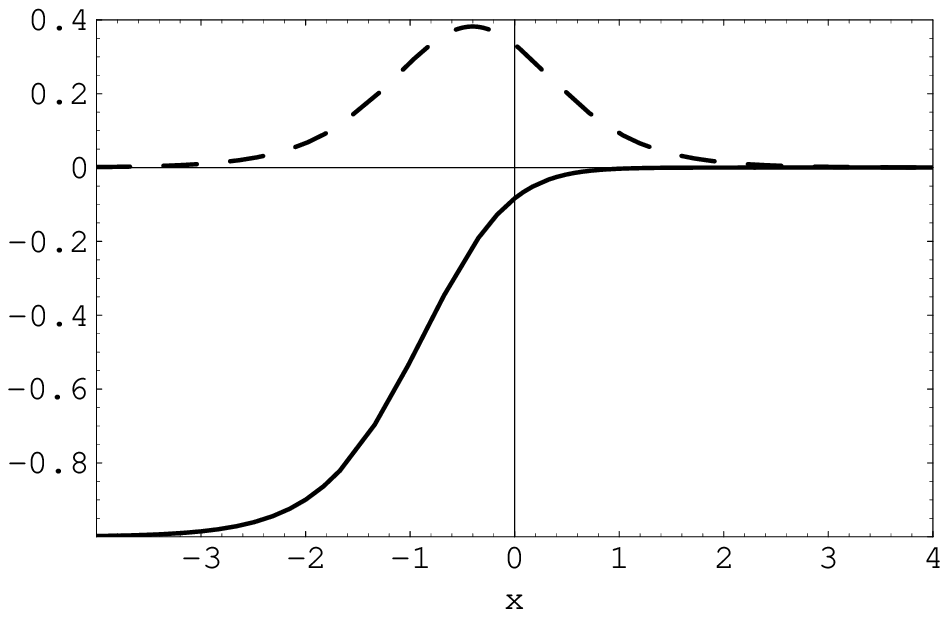}
\end{minipage}
\end{center}
\caption{Typical type A kink profile (for $\lambda=\mu$ ). The dotted line
corresponds to the field $\chi_+(x)$ and the solid line to the field $%
\phi_+(x)$. Both were calculated for $c_0=-3$.}
\label{fig:fig2}
\end{figure}

\begin{figure}[tbp]
\begin{center}
\begin{minipage}{20\linewidth}
\epsfig{file=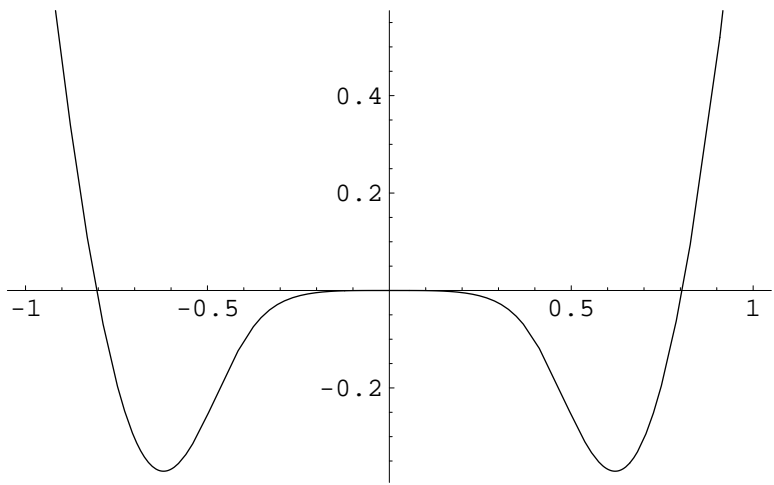}
\end{minipage}
\end{center}
\caption{The function $\sigma$, for $M=4$ as a function of the field $\chi$,
as defined in Eq. (35).}
\label{fig:fig3}
\end{figure}

\end{document}